\documentclass{ws-procs9x6}

\usepackage{subfigure}

\begin{document}

\title{A multiscale autocorrelation function for anisotropy studies}

\author{M. Scuderi$^*$}
\address{Dipartimento di Fisica e Astronomia, Universit\'a degli Studi di Catania  
\& Istituto Nazionale di Fisica Nucleare, Sez. di Catania, Via S. Sofia 64\\
Catania, 95123, Italy\\
$^*$E-mail: mario.scuderi@ct.infn.it\\
http://www.dfa.unict.it/}

\author{M. De Domenico}

\address{Laboratorio sui Sistemi Complessi, Scuola Superiore di Catania, Via Valdisavoia 9,\\
Catania, 95123, Italy\\}
\address{Istituto Nazionale di Fisica Nucleare - Sez. di Catania, Via S. Sofia 64\\
Catania, 95123, Italy\\
E-mail: manlio.dedomenico@ct.infn.it\\}

\author{A. Insolia}
\address{Dipartimento di Fisica e Astronomia, Universit\'a degli Studi di Catania  
\& Istituto Nazionale di Fisica Nucleare - Sez. di Catania, Via S. Sofia 64\\
Catania, 95123, Italy\\
E-mail: antonio.insolia@ct.infn.it\\}

\author{H. Lyberis}
\address{ 
IPN Orsay CNRS/IN2P3 and Universit\'e Paris Sud, Rue Georges Clemenceau 15\\
Orsay, 91406, France\\}
\address{Universit\'e Paris VII Denis Diderot, Paris 7, Paris Cedex 13\\
Paris, 75205, France\\}
\address{
Dipartimento di Fisica Generale, Universit\'a di Torino, Via Pietro Giuria 1\\
Torino, 10125, Italy\\
E-mail: lyberis@ipno.in2p3.fr\\}

\begin{abstract}
In recent years many procedures have been proposed to check the anisotropy of a dataset. 
We present a new simple procedure, based on a scale dependent approach, to detect anisotropy signatures in a given distribution with particular attention to small dataset.
The method provides a good discrimination power for both large and small datasets, even in presence of strong contaminating isotropic background. 
We present some applications to simulated datasets of events to investigate statistical features of the method and present and inspect its behavior under 
both the null or the alternative hypothesis.
\end{abstract}


\bodymatter

\section{Introduction}

In many field involving data analysis, the search for anisotropy has played a crucial role.
Many experimental data such as cosmological, astrophysical, atmospherical or geophysical ones require the search for clustering of objects and the 
measure of deviation from isotropy of a given angular distribution.
During the last decades, many estimators, namely correlation function $[$\refcite{Davis-1983, Szalay-1993, Hamilton-1993}$]$, have been proposed.
These methods apply to angular coordinates of objects as well to distributions of arrival directions of events.

Different estimators have been defined in the common anisotropy analysis, by Peebles, Davis-Peebles, Landy-Szalay and Hamilton $[$\refcite{Davis-1983, Szalay-1993, Hamilton-1993}$]$.
Recently, new estimators have been introduced to study the anisotropy signature of a given direction distributions: the modified two-point Rayleigh $[$\refcite{ave20092pt+}$]$, and shape-strength method derived from a principal component analysis of triplets of events $[$\refcite{Hague09}$]$.

Within the present work, we present a new fast and simple method for anisotropy analysis, which makes use of a multiscale approach and depends on one parameter only. The main advantage of our estimator is the possibility to analytically treat the results: the analytical approach drastically reduces computation time and makes available the possibility of applications to very large datasets. 
We test the method on several simulated isotropic and anisotropic arrival direction distributions and perform an extensive analysis of its statistical features under both the null and the alternative hypotheses.


\section{MAF: Multiscale Autocorrelation Function}\label{Scale}

Let $\mathcal{S}$ be a region of a spherical surface defining a sky and let $P_{i}(\phi,\theta)$ ($i=1,2,...,n$) be a set of $n$ points on $\mathcal{S}$. The sky $\mathcal{S}$ is partitioned within a grid of $N$ equal-area (and almost-equal shape) disjoint boxes $\mathcal{B}_{k}$ ($k=1,2,...,N$) as described in Ref. $[$\refcite{stokes2004using}$]$. Each box $\mathcal{B}_{k}$ covers the solid angle 
\begin{eqnarray}
\Omega_{k} = \frac{1}{N}\int_{\theta_{\text{min}}}^{\theta_{\text{max}}}\int_{\phi_{\text{min}}}^{\phi_{\text{max}}}d\cos\theta d\phi = 2\pi(1-\cos\Theta)\nonumber
\end{eqnarray}
where $2\Theta$ is the apex angle of a cone covering the same solid angle.
Let $\Omega$ be the solid angle covered by $\mathcal{S}$: $N,\Theta$ and $\Omega$ are deeply related quantities that define a scale.

The function $A(\Theta)$ that quantifies the deviation of data from an isotropic distribution at the scale $\Theta$, is chosen to be the Kullback-Leibler divergence $[$\refcite{kullback1951information, kullback1987kullback}$]$
\begin{eqnarray}
\label{def-A}
A(\Theta) = \mathcal{D}_{\text{KL}}( \psi(\Theta) || \overline{\psi}(\Theta)) = \sum_{k=1}^{N} \psi_{k}(\Theta)\log\frac{\psi_{k}(\Theta)}{\overline{\psi}_{k}(\Theta)}
\end{eqnarray}
where $\psi_{k}(\Theta)$ is the fraction of points in the dataset that fall into the box $\mathcal{B}_{k}$ and $\overline{\psi}_{k}(\Theta)$, generally a function of the domain meshing, is the expected number of points  on $\mathcal{B}_{k}$ from an isotropic distribution of $n$ points on $\mathcal{S}$. The Kullback-Leibler divergence is an information theoretic measure that quantifies the error in selecting the density $\overline{\psi}(\Theta)$ to approximate the density $\psi(\Theta)$ and it is strictly connected to maximum likelihood estimation (see Appendix A in Ref. $[$\refcite{JCAP2011}$]$). 
 
It is straightforward to show that $A(\Theta)$ is minimum for an isotropic distribution of points or, in general, when $\psi(\Theta)\sim \overline{\psi}(\Theta)$.
If $A_{\text{data}}(\Theta)$ and $A_{\text{iso}}(\Theta)$ refer to the data and to an isotropic realization with the same number of points respectively, we define \emph{multiscale autocorrelation function} (MAF) the estimator
\begin{eqnarray}
\label{def-s}
s(\Theta)=\frac{\left|A_{\text{data}}(\Theta)-\left\langle A_{\text{iso}}(\Theta) \right\rangle\right|}{\sigma_{A_{\text{iso}}}(\Theta)}
\end{eqnarray}
where $\left\langle A_{\text{iso}}(\Theta) \right\rangle$ and $\sigma_{A_{\text{iso}}}(\Theta)$ are the sample mean and the sample standard deviation, estimated from several isotropic realizations of the data. If $\mathcal{H}_{0}$ denotes the null hypothesis of an underlying isotropic distribution for the data, the chance probability at the angular scale $\Theta$, properly penalized because of the scan on $\Theta$, is the probability 
\begin{eqnarray}
\label{def-p}
p(\Theta) = \text{Pr}( s_{\text{iso}}(\Theta') \geq s_{\text{data}}(\Theta) | \mathcal{H}_{0}, \forall \Theta'\in\mathcal{P})
\end{eqnarray}
obtained from the fraction of null models giving a multiscale autocorrelation, at any angular scale $\Theta'$ in the parameter space $\mathcal{P}$, greater or equal than that of data at the scale $\Theta$\footnote{The null hypothesis is eventually rejected in favor of the alternative $\mathcal{H}_{1}=\lnot \mathcal{H}_{0}$ $-$ being $\lnot$ the negation operator $-$ at the angular scale $\Theta$, with probability $1-p(\Theta)$.}.
Under the null hypothesis $\mathcal{H}_{0}$, the estimator $s(\Theta)$ follows an half-gaussian distribution, independently on the value of the angular scale $\Theta$ and on the number of events on $\mathcal{S}$, as it will be successively shown in the text.


\section{Dynamical Counting} \label{sec-dyncount}

\begin{figure}[!t]  
\centering
  	\subfigure[\, Static counting]{ 
	  \includegraphics[width=3.5cm]{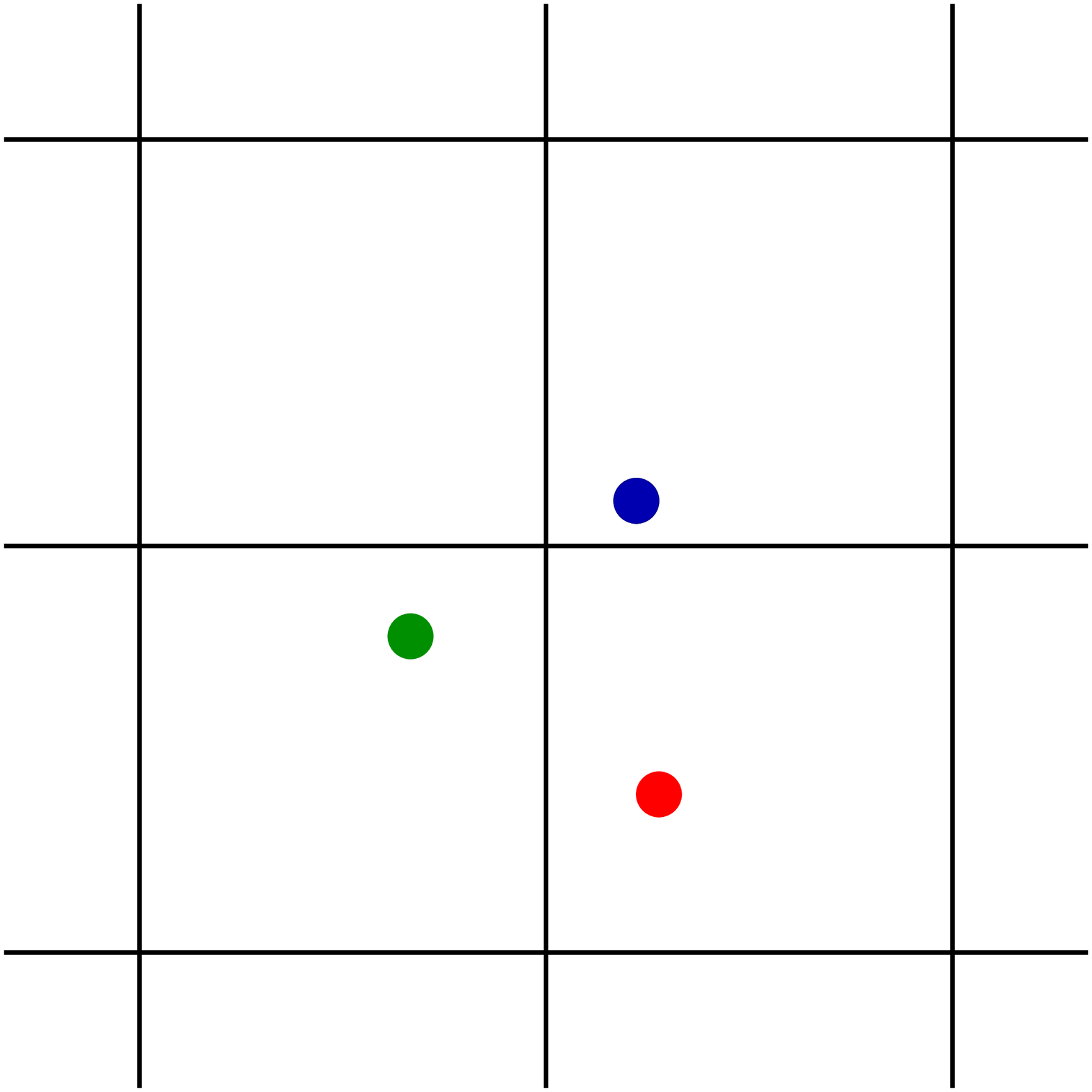}}
	\subfigure[\, Dynamic counting]{ 
	  \includegraphics[width=3.5cm]{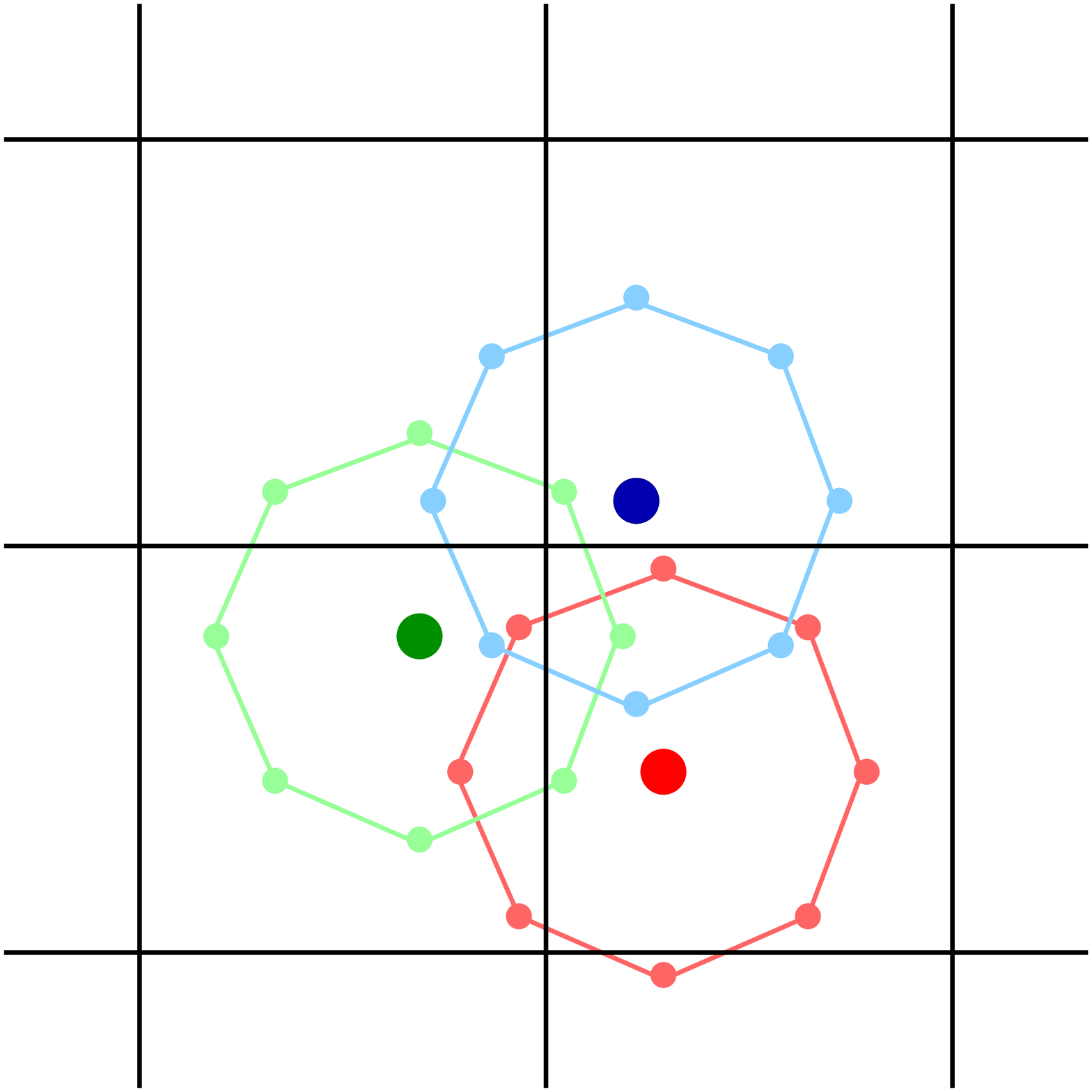}}
	  \caption{A cluster of three points falling into different boxes.}
\label{points}
\end{figure}

The simplest definition of the counting algorithm involves the fixed grid introduced in Ref. $[$\refcite{stokes2004using}$]$. In some cases, such a \emph{static counting} approach could not reveal an existing cluster. For instance, Figure\,\ref{points}(a) shows a typical scenario where some points of a given triplet fall into different cells. Indeed, the fixed grid may cut a cluster of points within one or more edges.

The smoothing, adopted in our study, deals with a new counting procedure for the estimation of the density $\psi_{k}(\Theta)$. 

From the starting distribution of points, we define a new distribution defined as follows:
for each point $P_{i}$, we consider a set of 8 points distributed around $P_{i}$ as described in Figure\,\ref{points}(b), being the distance between each of this points and $P_{i}$ equal to $\frac{\Theta}{2}$. Within this framework, from each point $P_{i}$ we define a set of 9 points ($P_{i}$ and the new 8 points), and assign to each point the weight $\frac{1}{9}$.
 Finally, we follow the standard procedure described in Section \ref{Scale} by using the weighted distribution of points\footnote{In general, in the weighting procedure, we have to take into account many factors, as for example the exposure of the experiment. See Ref. $[$\refcite{JCAP2011}$]$ for more details.}. 

Our numerical studies show that such a \emph{dynamical counting} approach gives more information on the amount of clustering in the data. 

\begin{figure}[!t]  
\centering
  	\subfigure[\, Clustered events]{ 
	  \includegraphics[width=4cm]{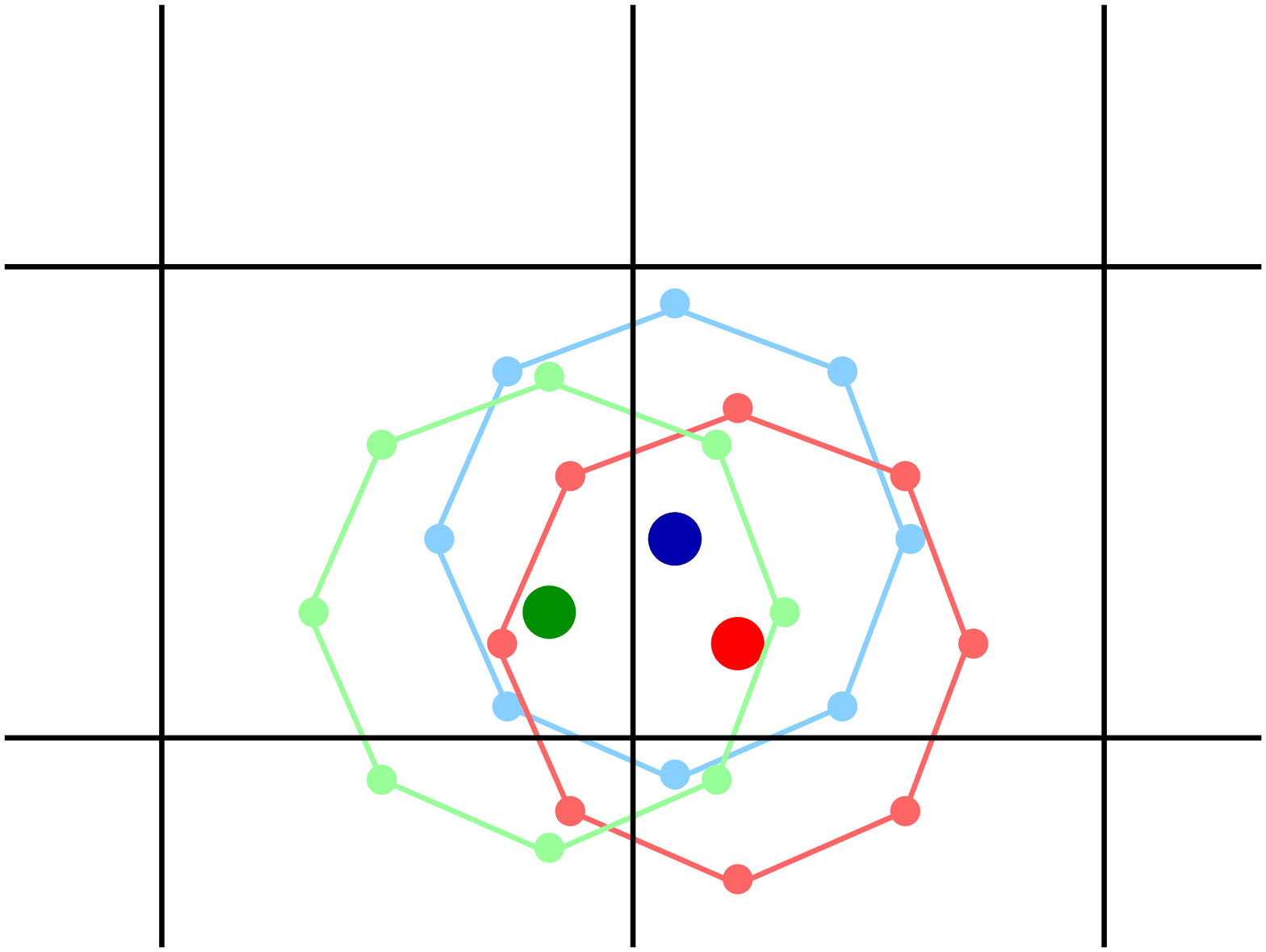}}
	\subfigure[\, Unclustered events]{ 
	  \includegraphics[width=4cm]{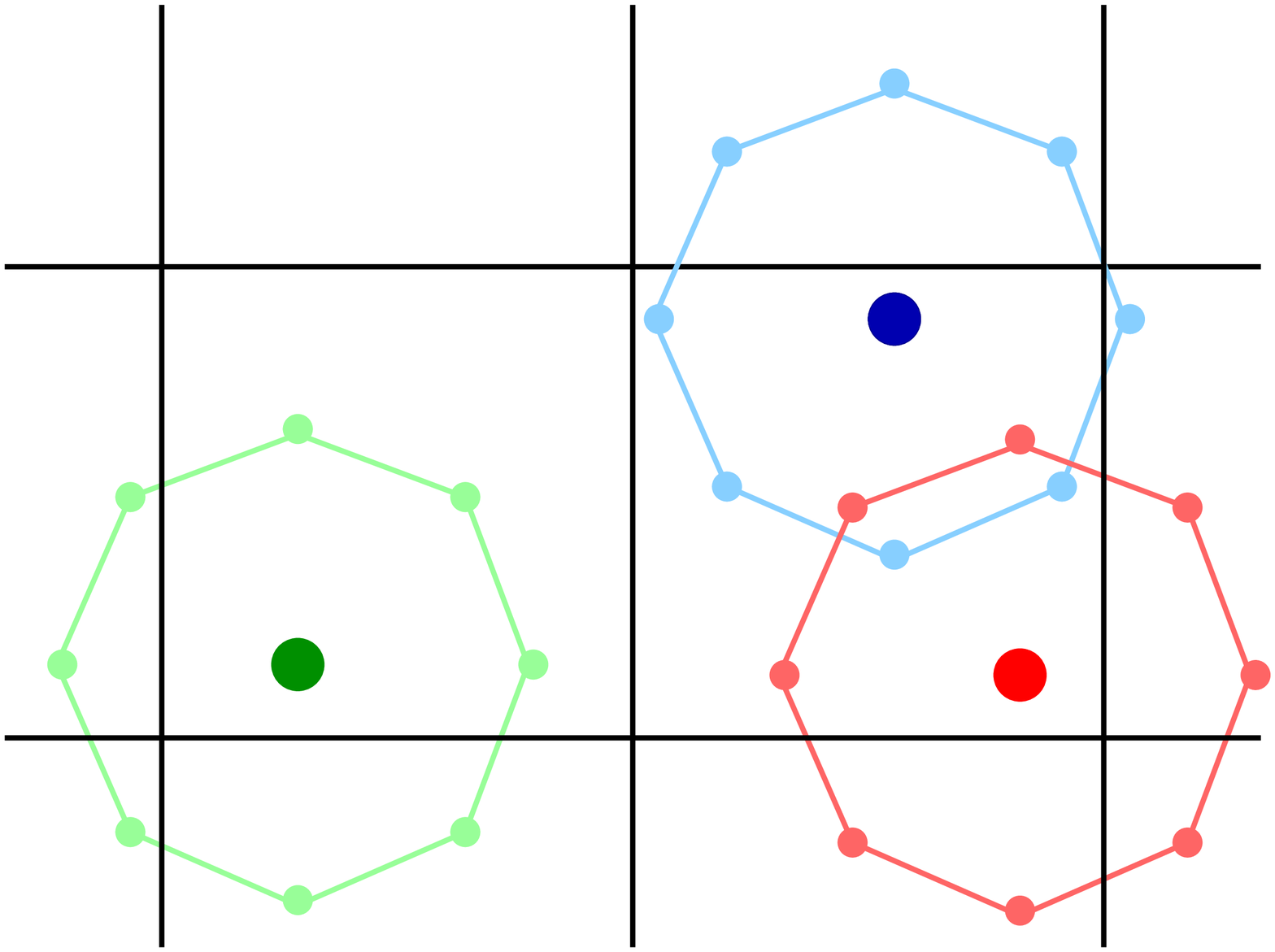}}
	  \caption{a) Three clustered points: the weighted points are concentrated in two adjacent boxes. b) Three unclustered points: the weighted points are distributed on the neighbor cells.}
\label{cluster}
\end{figure}

The main difference between the static and the dynamical counting lies in the value of the estimator when the procedure is applied to Monte Carlo realizations of the sky. For instance, let us consider the Figure \ref{cluster}, where we show a clustered (Figure \ref{cluster}(a)) and an unclustered (Figure \ref{cluster}(b)) set of points. The static counting is not able to recover the differences between the two configurations. Conversely, if the dynamical counting is applied, the \emph{extended} points in Figure \ref{cluster}(a) are concentrated mainly in two adjacent boxes while in Figure \ref{cluster}(b) they are distributed on the neighbor cells, this lead to two different $\psi(\Theta)$. Monte Carlo skies producing the same clustered configuration shown in Figure \ref{cluster}(a), and of consequence the same weight distribution, are not frequently expected: in this case, the value of $s(\Theta)$ should be greater than that one estimated from the static method. The direct consequence of a greater value of the estimator $s(\Theta)$ is a lower chance probability and the main advantage of using the dynamical counting, instead of the static one, should be the lowest penalization of $s(\Theta)$ only if an anisotropy signal is really present.

 In order to illustrate the importance of dynamical counting in the anisotropy signal detection, we have generated 5000 isotropic and anisotropic skies of 100 events each. In each anisotropic sky, 60\% of events are normally distributed, with dispersion $\rho$, around 10 random sources and 40\% of events are isotropically distributed. For each angular scale $\Theta$, we have estimated the average value of $s(\Theta)$ with the static and the dynamical counting, separately. Results are shown in Figure \ref{staticvsdyn} for $\rho=5^{\circ}$ (a), $\rho=10^{\circ}$ (b), $\rho=20^{\circ}$ (c) and for the isotropic map (d). In the case of the anisotropic skies, the dynamical counting provides a greater estimation of $s(\Theta)$ than the static counting, leading to a smaller estimation of the corresponding chance probability and improving the signal-to-noise ratio.

\section{Interpretation of MAF}


\begin{figure}[!t]  
\centering
	  \includegraphics[width=11cm]{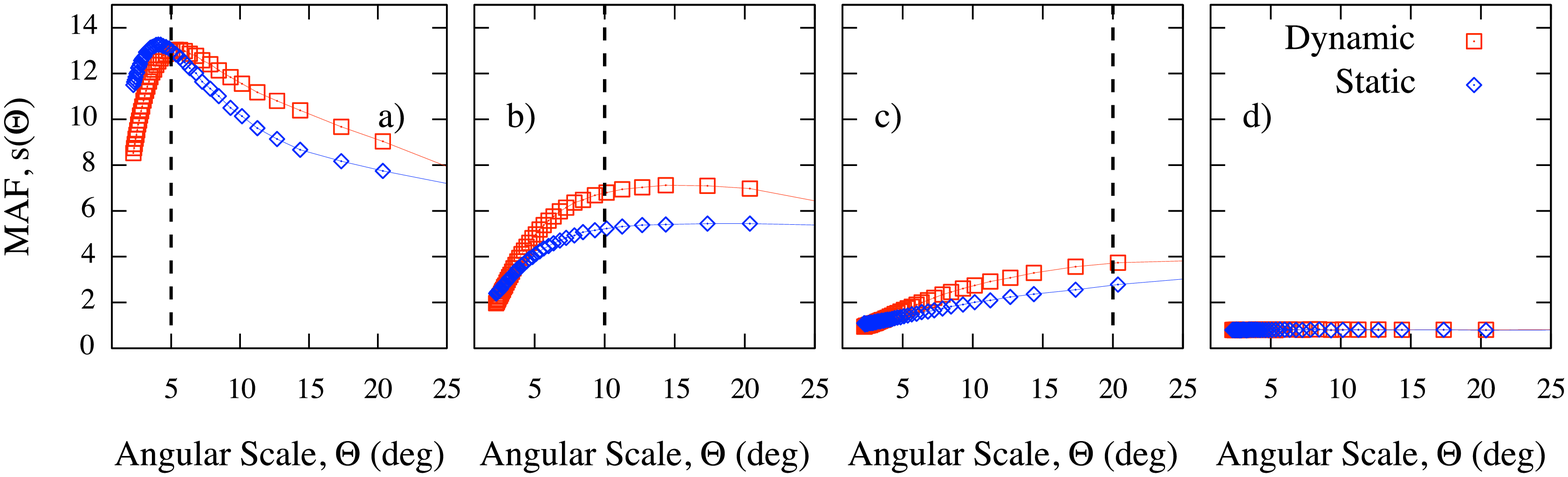}
	  \caption{MAF: average $s(\Theta)$ (solid line) estimated from $5000$ isotropic and anisotropic skies of 100 events each. In each anisotropic sky, 60\% of events are normally distributed, with dispersion $\rho$, around 10 random sources and 40\% of events are isotropically distributed. The dashed line indicates the value of the dispersion adopted to generate the corresponding mock map: a) 5, b) 10 and c) 20 degrees; d) isotropic map.}
\label{staticvsdyn}
	  \includegraphics[width=11cm]{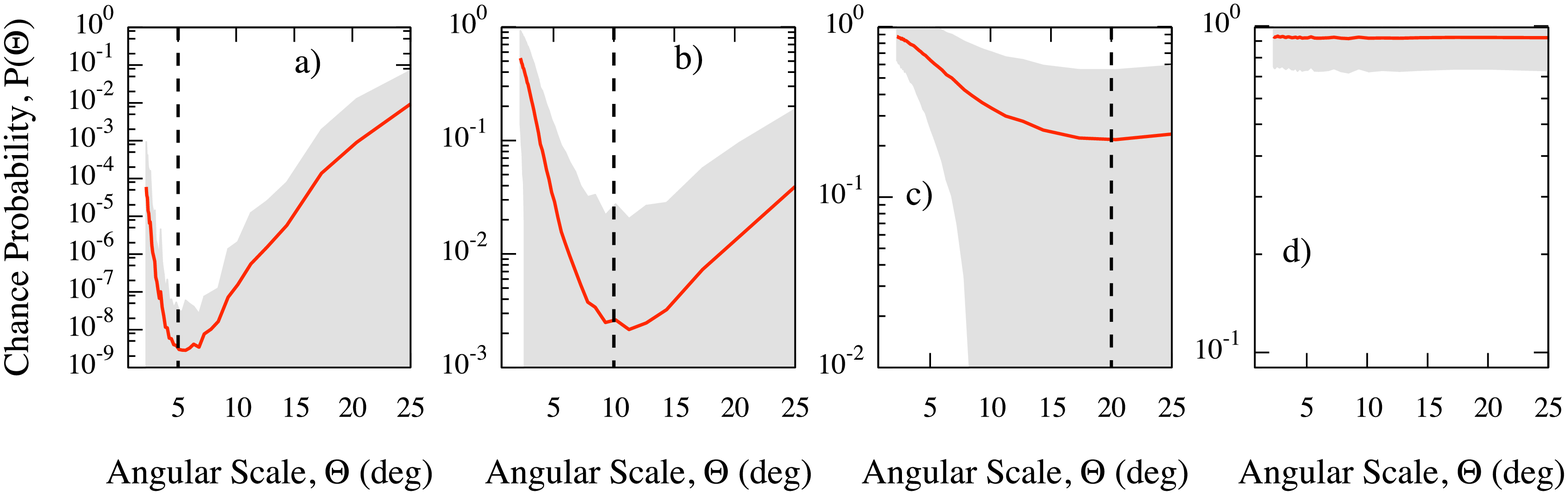}
	  \caption{MAF: average chance probability (solid line), with 68\% region around the mean value, estimated from isotropic and anisotropic skies generated as explained in Figure \ref{staticvsdyn}. Dynamical counting is used. The dashed line indicates the value of the dispersion adopted to generate the corresponding mock map: a) 5, b) 10 and c) 20 degrees; d) isotropic map.}
\label{AngScaleMock2}
\end{figure}

In the specific case of MAF, the angular scale $\Theta^{\star}$, where the significance is minimum, turns to be the significative {\itshape{clustering scale}}: it is the scale at which occurs a greater accumulation of points respect to that one occuring by chance, with no regard for a particular configuration of points, e.g. doublets or triplets. To illustrate better the clustering scale detection feature of MAF, we have generated 5000 isotropic and anisotropic skies of 100 events each, as previously described at the end of Section \ref{sec-dyncount}. Figure\,\ref{AngScaleMock2} shows the average chance probability, with 68\% region around the mean value, versus the angular scale for three values of the dispersion, namely $\rho=5^{\circ}$ (a), $\rho=10^{\circ}$ (b), $\rho=20^{\circ}$ (c), and for the isotropic map (d). As expected, chance probability is close to one and nearly flat in the case of the isotropic map, because all clustering scales are equally likely. Conversely, for all anisotropic maps, the average chance probability gets a minimum around the corresponding value of $\rho$. Thus, our estimator is able to recover the most significant clustering scale. It should be remarked that when the $20^{\circ}$ dispersion is used, the angular scale of the minimum is less obvious because of the large fluctuations due to the isotropic contamination. Finally, it is worth noticing that we have observed that the curve around the value of $\rho$ gets narrower by increasing the number of events.


\section{Statistical analysis of MAF}

We have investigated the statistical features of MAF by inspecting its behavior under both the null or the alternative hypothesis (see Ref. $[$\refcite{JCAP2011}$]$ for further information). In particular, we have estimated the significance $\alpha$ (or Type I error), i.e. the probability to wrongly reject the null hypothesis when it is actually true, and the power $1-\beta$ (where $\beta$ is known as Type II error), i.e. the probability to accept the alternative hypothesis when it is in fact true. In the following we will adopt the dynamical counting previously discussed.

\vspace{0.25truecm}\hspace{0.1truecm}{\bfseries{Null hypothesis.}} We generate isotropic maps of $10^{5}$ skies, by varying the number of events from 20 to 500 and estimate the MAF for several values of the angular scale $\Theta$. By choosing: 
\begin{eqnarray}
\tilde{p}(\Theta^{\star})=\arg\min_{\Theta}p(\Theta)\nonumber
\end{eqnarray}

We find an excellent flat distribution of probabilities $\tilde{p}(\Theta^{\star})$, as expected for analyses under the null hypothesis $\mathcal{H}_{0}$. In other words, MAF is not biased against $\mathcal{H}_{0}$, as required for good statistical estimators.

Because of the definition in Eq. (\ref{def-A}) and of the central limit theorem, a Gaussian distribution is expected for the function $A(\Theta)$, and of consequence, the half-normal distribution 
\begin{eqnarray}
\mathcal{G}_{1/2}[s(\Theta)] = \frac{2}{\sqrt{2\pi}\sigma(\Theta)}e^{-\frac{s^{2}(\Theta)}{2\sigma^{2}(\Theta)} }
\end{eqnarray}
for $\sigma(\Theta)=1$, is expected for the estimator $s(\Theta)$ defined as in Eq. (\ref{def-s}), being normalized to zero mean and unitary variance. It follows that the (unpenalized) probability to obtain by chance a value of the MAF, greater or equal than a given value $s_{0}$, is just $1-\text{erf}(\frac{s_{0}}{\sqrt{2}})$, being erf the standard error function, independently of the angular scale $\Theta$.

Although this nice feature of the MAF estimator it is more important to identify the distribution of the penalized probability $p(\Theta)$, because of the required penalization due to the scan over the parameter $\Theta$.
For this reason, the distribution of $s_{\text{max}}=\max\{s(\Theta)\}$ is of interest for applications.

Our numerical studies show that such a distribution exists and it corresponds to one of the limiting densities in the extreme value theory (see Appendix B in Ref. $[$\refcite{JCAP2011}$]$). 

\begin{figure}[!t]
	\centering
	  \includegraphics[width=8cm]{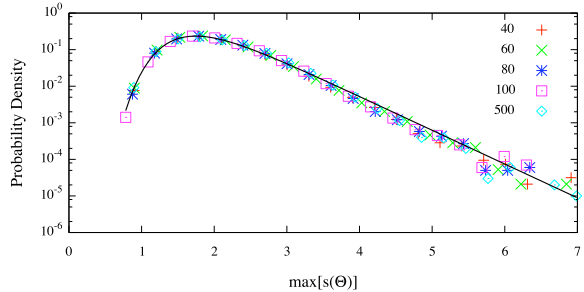}
	\caption{MAF: distributions of $\max\{s(\Theta)\}$ for $n=40, 60, 80, 100$ and $500$ events. Solid line correspond to the least-square fit of the Gumbel density with parameters $\mu=1.743  \pm 0.002$ and $\sigma=0.470   \pm 0.002$ ($\chi^{2}/\text{ndf}=1.1\times 10^{-5}$).}
\label{smax-gumbel}
\end{figure}

In Figure \ref{smax-gumbel} are shown the probability densities of $s_{\text{max}}$ for $n=40, 60, 80, 100$ and $500$ events: independently on $n$, each density is in excellent agreement with the Gumbel distribution of extreme values, for the parameters $\mu= 1.743  \pm 0.002$ and $\sigma= 0.470   \pm 0.002$. Such values correspond to the mean and to the standard deviation of the distribution, $\tilde{\mu}\approx 2.00$ and $\tilde{\sigma}\approx 0.59$, respectively. It follow that the probability to obtain a maximum value of $s(\Theta)$, at any angular scale $\Theta$, greater or equal than a given value $\max\{s(\Theta)\}$ is
\begin{eqnarray}
p(\max\{s(\Theta)\})=1-\exp[ -\exp(\frac{\max\{s(\Theta)\}-\mu}{\sigma})],\nonumber
\end{eqnarray}

\vspace{0.25truecm}\hspace{0.1truecm}{\bfseries{Alternative hypothesis.}} In order to investigate the behavior of MAF under the alternative hypothesis, we have generated anisotropic maps of $10^{4}$ skies, by varying the number of events from 20 to 100. 


In order to estimate the power of MAF, we build reasonable anisotropic maps of CR events reflecting in part the real-world scenario, keeping in mind that our purpose is to build an anisotropic set of events for statistical analysis and not to generate events mimicking real datasets with the best available approximation (see Ref. $[$\refcite{JCAP2011}$]$ for further information).



\begin{figure}[!h]
	\centering
	  \includegraphics[width=8cm]{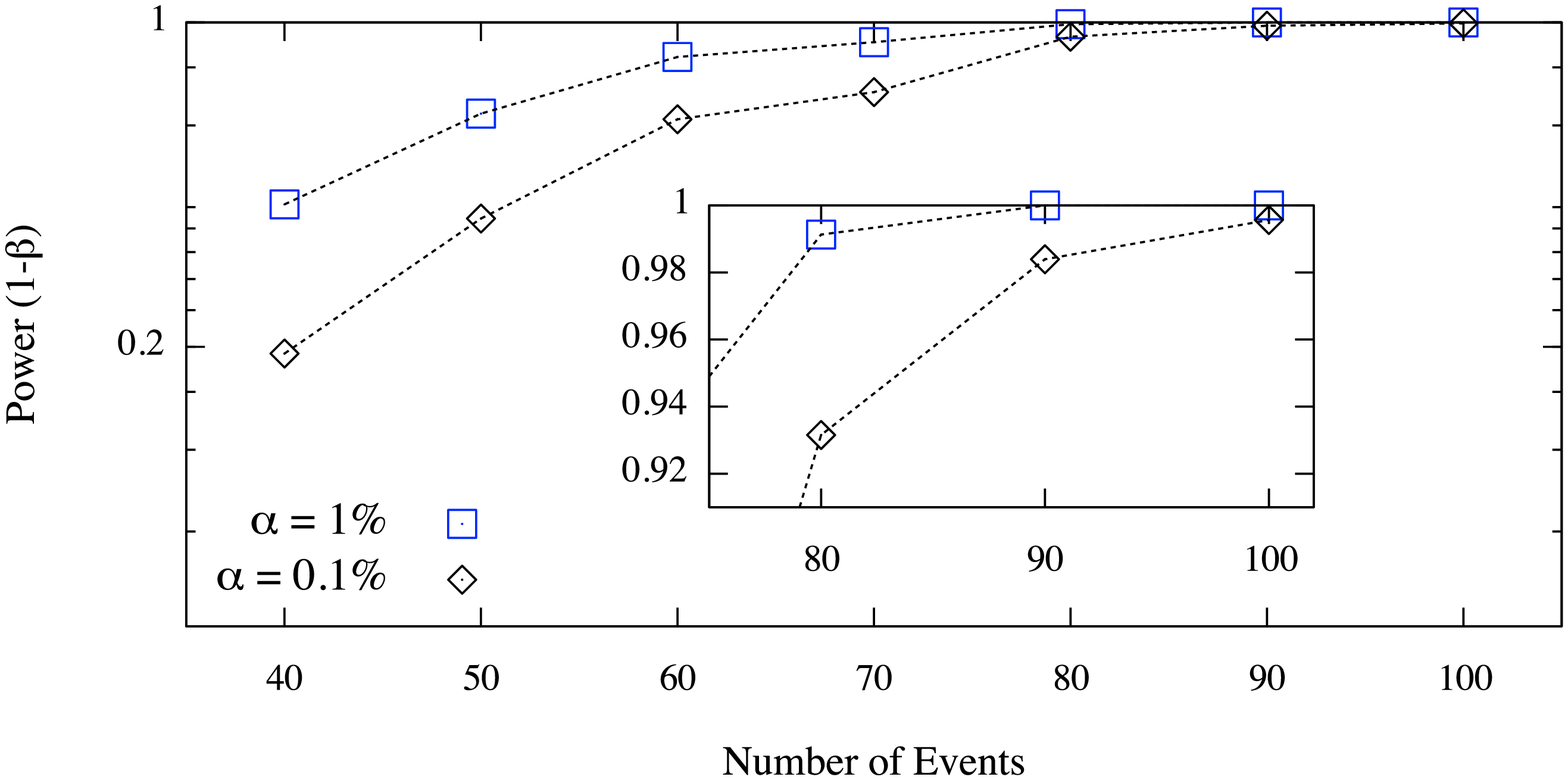}
	\caption{MAF power vs. the number of events sampled from anisotropic mock maps generated as described in the text, for values of the significance corresponding to $\alpha=0.1\%$ and $\alpha=1\%$.}
\label{maf-power}
\end{figure}

A sky is labelled as \emph{anisotropic} if, for a fixed value of the significance $\alpha$, the penalized chance probability as defined in Eq. (\ref{def-p}) is lesser or equal than $\alpha$, i.e. if the condition 
\begin{eqnarray}
\tilde{p}(\Theta^{\star})=\arg\min_{\Theta} p(\Theta)\leq \alpha\nonumber
\end{eqnarray}
holds for some angular scale $\Theta^{\star}$. In Figure \ref{maf-power} is shown the power for two values of the significance threshold, namely $\alpha=0.1\%$ and $\alpha=1\%$, estimated through the analytical approach. For applications, a power of 90\% is generally required: under this threshold the method could miss to detect an existing anisotropy signal. In the case of the MAF, and for the considered anisotropic mock map, the power increases with the number of events $n$ and it is able to detect the anisotropic signal for $n\geq 60$, with significance $\alpha=1\%$. However, by decreasing the significance for the statistical test, the power requires a greater number of events to reach the 90\% threshold, as expected.



\section{Conclusion}

We introduced a new statistical test, based on a multiscale approach, for detecting an anisotropy signal in the arrival direction distribution of UHECR. We showed that our procedure is suitable for the analysis of both small and large datasets of events. 

The advantages of this new approach over other methods are multiples. First, the method allows an analytical description of quantities involved in the estimation of the amount of anisotropy signal in the data, avoiding thousands of Monte Carlo realizations needed for the penalizing procedure of results. Second, the method allows the detection of a physical observable, namely the clustering scale. Third, the method is unbiased against the null hypothesis and it provides a high discrimination power even in presence of strong contaminating isotropic background, for both small and large datasets.

This method is suitable for the detection of the anisotropy signal in each data set involving a distribution of angular coordinates on the sphere, and it can be adapted to non-spherical spaces by properly redefining the dynamical counting algorithm.

\newpage

\bibliographystyle{ws-procs9x6}
\bibliography{ws-procs9x6}

\end{document}